# Exact Solutions of the Duffin-Kemmer-Petiau Equation for the Deformed Hulthen Potential

Fevziye Yaşuk, Cüneyt Berkdemir, Ayşe Berkdemir and Coşkun Önem [a]

*Department of Physics, Faculty of Arts and Sciences, Erciyes University, , Kayseri, Turkey*

## Abstract

Using the Nikiforov-Uvarov method, an application of the relativistic Duffin-Kemmer-Petiau equation in the presence of a deformed Hulthen potential is presented for spin zero particles. We derived the first-order coupled differential radial equations which enable the energy eigenvalues as well as the full wavefunctions to be evaluated by using of the Nikiforov-Uvarov method that can be written in terms of the hypergeometric polynomials.

Corresponding author:
[a] Coşkun Önem, e-mail: onem@erciyes.edu.tr

# 1. Introduction

Great deals of important problems in relativistic quantum mechanics as well as non-relativistic quantum mechanics have been solved by using Nikiforov–Uvarov method (NU) [1,2,3]. This method depends on the solution of second order linear differential equations whose solutions leads to hypergeometric type equations. Especially, exact energy levels of the exponential-type molecular potentials such as Hulthen potential [4] have been obtained analytically by making use of the NU-method in the Klein-Gordon equation [5]. Using this method gives us opportunity of obtaining both the energy eigenvalues and eigenfunctions systematically.

There has been a recent revival of interest in the Duffin–Kemmer–Petiau (DKP) equation and its relevance to some problems in nuclear and particle physics. Recently the first-order relativistic DKP equation has been used to study the interactions of spinless mesons with nuclei [6]. The problem of a relativistic particle with arbitrary spin has had a long history. The first-order relativistic wave equations for arbitrary spin were first investigated by Lubański, Madhavarao and Bhabha and have come to be known as Bhabha wave equations. The simplest special cases of these equations reduce to the Dirac and DKP equations for spin-1/2 and spins 0 and 1, respectively [7]. Following the success of the Dirac equation [8] in relativistically describing particles of spin-1/2, the search began for similar first-order wave equations for spin-0 and spin-1 particles. Apart from spin one-half, none of the other spins obeys a single relativistic wave equation [9]. For instance, it was generally believed that for spins zero and one, the second-order Klein-Gordon [10,11] and the second-order Proca [12] equations were unique. But, after that the Dirac-like first order relativistic Duffin-Kemmer-Petiau equations were found and it can define both spin-zero and spin-one particles [13,14,15].

The goal of this study is to obtain exact solutions of Duffin-Kemmer-Petiau equation in the presence of Hulthen potential for spin zero representation by using Nikiforov-Uvarov method. Therefore, we compare the eigenstates of DKP and KG equations with Hulthen potential and test the equivalence of these equations for spin zero particles. Firstly, we talk about Duffin-Kemmer-Petiau formalism briefly and derive second order radial differential equations with general vector potential. After that, NU method is given in terms of hypergeometric polynomials. Finally, deformed Hulthen potential is substituted in DKP equation and then energy eigenvalues and eigenfunctions are obtained.

# 2. Duffin-Kemmer-Petiau Formalizm

In this section, the Duffin-Kemmer-Petiau formalism is given briefly. Generally the first order relativistic Duffin–Kemmer–Petiau equation for a free spin zero or spin one particle of mass m is

$$(i\beta^\mu \partial_\mu - m)\psi_{DKP} = 0 \qquad (1)$$

where $\beta^\mu$ ($\mu = 0, 1, 2, 3$) matrices satisfy the commutation relation

$$\beta^\mu \beta^\nu \beta^\lambda + \beta^\lambda \beta^\nu \beta^\mu = g^{\mu\nu}\beta^\lambda + g^{\nu\lambda}\beta^\mu \qquad (2)$$

which defines the so-called Duffin–Kemmer–Petiau (DKP) algebra. The algebra generated by the 4 $\beta^N$'s has three irreducible representations: a ten dimensional one that is related to S=1, a five dimensional one relevant for S=0 (spinless particles) and a one dimensional one which is trivial.

In the spin-0 representation, $\beta^\mu$ are $5 \times 5$ matrices defined as ($i = 1,2,3$)

$$\beta^0 = \begin{pmatrix} \theta & \tilde{0} \\ \bar{0}_T & 0 \end{pmatrix}, \quad \beta^i = \begin{pmatrix} \tilde{0} & \rho^i \\ -\rho^i_T & 0 \end{pmatrix} \tag{3}$$

with $\tilde{0}$, $\bar{0}$, $0$ as $2\times 2$, $2\times 3$, $3\times 3$ zero matrices, respectively, and

$$\theta = \begin{pmatrix} 0 & 1 \\ 1 & 0 \end{pmatrix}, \quad \rho^1 = \begin{pmatrix} -1 & 0 & 0 \\ 0 & 0 & 0 \end{pmatrix}, \quad \rho^2 = \begin{pmatrix} 0 & -1 & 0 \\ 0 & 0 & 0 \end{pmatrix}, \quad \rho^3 = \begin{pmatrix} 0 & 0 & -1 \\ 0 & 0 & 0 \end{pmatrix} \tag{4}$$

For spin one particles, $\beta^\mu$ are $10\times 10$ matrices given by

$$\beta^0 = \begin{pmatrix} 0 & \bar{0} & \bar{0} & \bar{0} \\ \bar{0}^T & 0 & \mathbf{I} & \mathbf{I} \\ \bar{0}^T & \mathbf{I} & 0 & 0 \\ \bar{0}^T & 0 & 0 & 0 \end{pmatrix}, \quad \beta^i = \begin{pmatrix} 0 & \bar{0} & e_i & \bar{0} \\ \bar{0}^T & 0 & 0 & -is_i \\ e_i^T & 0 & 0 & 0 \\ \bar{0}^T & -is_i & 0 & 0 \end{pmatrix} \tag{5}$$

where $s_i$ are the usual $3\times 3$ spin one matrices

$$\bar{0} = \begin{pmatrix} 0 & 0 & 0 \end{pmatrix}, \quad e_1 = \begin{pmatrix} 1 & 0 & 0 \end{pmatrix}, \quad e_2 = \begin{pmatrix} 0 & 1 & 0 \end{pmatrix}, \quad e_3 = \begin{pmatrix} 0 & 0 & 1 \end{pmatrix}. \tag{6}$$

$\mathbf{I}$ and $0$ are the identity and zero matrices, respectively. While the dynamical state $\psi_{DKP}$ is a five component spinor for spin zero particles, it has ten component spinors for $S=1$ particles.

The solution of the DKP equation for a particle in a central field needs consideration since earlier work [6]. It is convenient to recall some general properties of the solution of the DKP equation in a central interaction for spin zero particles. The central interaction consists of two parts: a Lorentz scalar $U_s$ and a time-like vector $U_v$ potential. The stationary states of the DKP particle in this case are determined by solving

$$\left(\beta.\mathbf{p}\, c + m\, c^2 + U_s + \beta^0\, U_v^0 \right)\psi(\mathbf{r}) = \beta^0 E \psi(\mathbf{r}) \tag{7}$$

In the spin zero representation, the five component DKP spinor

$$\psi(\mathbf{r}) = \begin{pmatrix} \psi_{upper} \\ i\psi_{lower} \end{pmatrix} \quad \text{with} \quad \psi_{upper} \equiv \begin{pmatrix} \phi \\ \varphi \end{pmatrix} \quad \text{and} \quad \psi_{lower} \equiv \begin{pmatrix} A_1 \\ A_2 \\ A_3 \end{pmatrix}. \tag{8}$$

so that for stationary states the DKP equation can be written as

$$\left(m\, c^2 + U_s\right)\phi = \left(E - U_v^0\right)\varphi + \hbar c \nabla.\mathbf{A} \tag{9}$$

$$\nabla \phi = \left(m\, c^2 + U_s\right)\mathbf{A} \tag{10}$$

$$\left(m\, c^2 + U_s\right)\varphi = \left(E - U_v^0\right)\phi \tag{11}$$

where $\mathbf{A}$ is the vector $(A_1, A_2, A_3)$.

The five-component wavefunction $\psi$ is simultaneously an eigenfunction of $J^2$ and $J_3$

$$J^2 \begin{pmatrix} \psi_{upper} \\ \psi_{lower} \end{pmatrix} = \begin{pmatrix} L^2 \psi_{upper} \\ (L+S)^2 \psi_{lower} \end{pmatrix} = J(J+1)\begin{pmatrix} \psi_{upper} \\ \psi_{lower} \end{pmatrix} \tag{12}$$

$$J_3 \begin{pmatrix} \psi_{upper} \\ \psi_{lower} \end{pmatrix} = \begin{pmatrix} L_3 \psi_{upper} \\ (L_3+s_3)\psi_{lower} \end{pmatrix} = M \begin{pmatrix} \psi_{upper} \\ \psi_{lower} \end{pmatrix} \tag{13}$$

where the total angular momentum $J = L + S$ which commutes with $\beta^0$, is a constant of the motion. The most general solution of (7) is

$$\psi_{JM}(r) = \begin{pmatrix} f_{nJ}(r)Y_{JM}(\Omega) \\ g_{nJ}(r)Y_{JM}(\Omega) \\ i\sum_{L} h_{nJL}(r)Y_{JL1}^{M}(\Omega) \end{pmatrix} \quad (14)$$

where $Y_{JM}(\Omega)$ are the spherical harmonics of order $J$, $Y_{JL1}^{M}(\Omega)$ are the normalized vector spherical harmonics and $f_{nJ}$, $g_{nJ}$ and $h_{nJL}$ are radial wave functions. Inserting $\psi_{JM}(r)$ as given in (14) into (9), (10), (11) by using the properties of vector spherical harmonics [16] one gets the following set of first-order coupled relativistic differential radial equations

$$\left(E - U_v^0\right)F(r) = \left(m c^2 + U_s\right)G(r) \quad (15)$$

$$\left(\frac{dF(r)}{dr} - \frac{J+1}{r}F(r)\right) = -\frac{1}{\alpha_J}\left(m c^2 + U_s\right)H_1(r) \quad (16)$$

$$\left(\frac{dF(r)}{dr} + \frac{J}{r}F(r)\right) = \frac{1}{\zeta_J}\left(m c^2 + U_s\right)H_{-1}(r) \quad (17)$$

$$-\alpha_J\left(\frac{dH_1(r)}{dr} + \frac{J+1}{r}H_1(r)\right) + \zeta\left(\frac{dH_{-1}(r)}{dr} - \frac{J}{r}H_{-1}(r)\right)$$
$$= \frac{1}{\hbar c}\left(\left(m c^2 + U_s\right)F(r) - \left(E - U_v^0\right)G(r)\right) \quad (18)$$

with the definition of $\alpha_J = \sqrt{(J+1)/(2J+1)}$, $\zeta = \sqrt{J/(2J+1)}$ and $f_{nJ}(r) = F(r)/r$, $g_{nJ} = G(r)/r$ and $h_{nJJ\pm 1} = H_{\pm 1}/r$.

## 3. Nikiforov-Uvarov Method

In this method, for a given real or complex potential, the Schrödinger equation in one dimension is reduced to a generalized equation of hypergeometric type with an appropriate $s = s(x)$ coordinate transformation and it can be written in the following form,

$$\psi''(s) + \frac{\tilde{\tau}(s)}{\sigma}\psi'(s) + \frac{\tilde{\sigma}(s)}{\sigma^2(s)}\psi(s) = 0 \quad (19)$$

where $\sigma(s)$ and $\tilde{\sigma}(s)$ are polynomials, at most second-degree, and $\tilde{\tau}(s)$ is a first-degree polynomial. Hence, from the (19) the Schrödinger equation and the Schrödinger-like equations can be solved by means of the special potentials or some quantum mechanics problems. To find particular solution of (19) by separation of variables, if one deals with the transformation

$$\psi(s) = \phi(s)y(s) \quad (20)$$

it reduces to an equation of hypergeometric type,

$$\sigma(s)y'' + \tau(s)y' + \lambda y = 0 \quad (21)$$

where $\phi(s)$ satisfies $\phi(s)'/\phi(s) = \pi(s)/\sigma(s)$. $y(s)$ is the hypergeometric type function whose polynomial solutions are given by Rodrigues relation

$$y_n(s) = \frac{B_n}{\rho(s)}\frac{d^n}{ds^n}\left[\sigma^n(s)\rho(s)\right], \quad (22)$$

where $B_n$ is a normalizing constant and the weight function $\rho$ must satisfy the condition

$$(\sigma\rho)' = \tau\rho. \tag{23}$$

The function $\pi$ and the parameter $\lambda$ required for this method are defined as

$$\pi = \frac{\sigma' - \tilde{\tau}}{2} \pm \sqrt{\left(\frac{\sigma' - \tilde{\tau}}{2}\right)^2 - \tilde{\sigma} + k\sigma} \tag{24}$$

and

$$\lambda = k + \pi'. \tag{25}$$

On the other hand, in order to find the value of $k$, the expression under the square root must be square of polynomial. Thus, a new eigenvalue equation for the Schrödinger equation becomes

$$\lambda = \lambda_n = -n\tau' - \frac{n(n-1)}{2}\sigma'', \quad (n = 0,1,2,...) \tag{26}$$

where

$$\tau(s) = \tilde{\tau}(s) + 2\pi(s), \tag{27}$$

and derivative of that is negative.

## 4. The Deformed Hulthen potential for DKP equation

The Hulthen potential [17] is one of the important molecule potentials with an exponential-type included a wide class of potentials in several branches of physics science. This potential is studied by many authors solving relativistic or non-relativistic differential-type equations [18,19,20] and its deformed shape is defined by

$$U_v^0 = -Ze^2\delta \frac{e^{-\delta r}}{(1 - qe^{-\delta r})} \tag{28}$$

where Z is a constant, $\delta$ and $q$ are the screening and deformation parameters, respectively. For small r compared to $1\delta$, the Hulthen potential approaches the Coulomb potential whereas for large r compared to $1\delta$, it approaches to zero exponentially. Therefore, $1\delta$ term can be thought as an infrared regulator for the Coulomb problem [21]. In addition, for $q = 1$ the deformed form reduces to the Hulthen potential form and for $q = -1$ to the standard Woods-Saxon potential.

For DKP equation, at the presence of vector potential and while scalar potential is zero, by choosing $\hbar = 1, c = 1$, the differential equations to be satisfied by the radial wavefunctions are

$$\left(E - U_v^0\right)F(r) = mG(r) \tag{29}$$

$$\left[\frac{dF(r)}{dr} - \frac{J+1}{r}F(r)\right] = -\frac{1}{\alpha_J}mH_1(r) \tag{30}$$

$$\left[\frac{dF(r)}{dr} + \frac{J}{r}F(r)\right] = \frac{1}{\zeta_J}mH_{-1}(r) \tag{31}$$

$$-\alpha_J\left(\frac{dH_1(r)}{dr} + \frac{J+1}{r}H_1(r)\right) + \zeta_J\left(\frac{dH_{-1}(r)}{dr} - \frac{J}{r}H_{-1}(r)\right)$$
$$= mF(r) - \left(E - U_v^0\right)G(r) \tag{32}$$

Eliminating $G(r)$, $H_1(r)$ and $H_{-1}(r)$ in terms of $F(r)$ the following second-order equation is obtained

$$\left[\frac{d^2}{dr^2} - \frac{J(J+1)}{r^2} + (E - U_v^0)^2 - m^2\right] F(r) = 0. \tag{33}$$

Equation (31) appears as the radial KG equation for a vector potential. Let us now apply the NU method to the problem of finding the eigenvalues and the eigenfunctions of the radial DKP equation for the deformed Hulthen potential choosing $J = 0$ (s-wave case). The deformed potential form given by (26) is substituted in the radial DKP equation

$$F_q(s)'' + \left[E^2 + 2V_0 E \frac{e^{-\delta r}}{1 - qe^{-\delta r}} + V_0^2 \frac{e^{-2\delta r}}{(1 - qe^{-\delta r})^2} - m^2\right] F_q(s) = 0 \tag{34}$$

and when a transformation of the form $s = e^{-\delta r}$ is performed together with the following dimensionless parameters

$$\varepsilon^2 = -\frac{1}{\delta^2}\left[E^2 - m^2\right] \geq 0 (E^2 \leq m^2), \quad \beta^2 = \frac{2V_0 E}{\delta^2}, \quad \gamma^2 = \frac{V_0^2}{\delta^2} \tag{35}$$

this equation is reduced to the generalized equation of hypergeometric type which is given by (17)

$$F_q(s)'' + \frac{1-sq}{s(1-sq)} F_q(s)' + \frac{1}{[s(1-sq)]^2}\left[s^2(\gamma^2 - q^2\varepsilon^2 - q\beta^2) + s(\beta^2 + 2q\varepsilon^2) - \varepsilon^2\right] F_q(s) = 0 \tag{36}$$

with $\tilde{\tau}(s) = 1 - qs$, $\sigma(s) = s(1-qs)$, $\tilde{\sigma}(s) = s^2(\gamma^2 - q^2\varepsilon^2 - q\beta^2) + s(\beta^2 + 2q\varepsilon^2) - \varepsilon^2$. According to the NU method the function $\pi(s)$ is obtained as

$$\pi(s) = -\frac{qs}{2} \pm \frac{1}{2}\sqrt{s^2\left[q^2 - 4(\gamma^2 - q^2\varepsilon^2 - q\beta^2) - 4qk\right] + 4s\left[k - (\beta^2 + 2q\varepsilon^2)\right] + 4\varepsilon^2} \tag{37}$$

where the parameter k is to be determined by the condition that the expression under the square root's discriminant is zero. One then obtains the following possible solutions for each k:

$$\pi(s) = -\frac{qs}{2} \pm \frac{1}{2}\begin{cases}[(2q\varepsilon - \alpha)s - 2\varepsilon] & \text{for} \quad k = \beta^2 - \alpha\varepsilon \\ [(2q\varepsilon - \alpha)s + 2\varepsilon] & \text{for} \quad k = \beta^2 + \alpha\varepsilon\end{cases} \tag{38}$$

where $\alpha = \sqrt{q^2 - 4\gamma^2}$. For the polynomial of $\tau(s) = \tilde{\tau}(s) + 2\pi(s)$ which has a negative derivative, we select $k = \beta^2 - \alpha\varepsilon$ and $\pi(s) = -\frac{qs}{2} - \frac{1}{2}[(2q\varepsilon - \alpha)s + 2\varepsilon]$. With this selection and $\lambda = k + \pi'$, $\tau(s)$ and $\lambda$ can be written as, respectively,

$$\tau(s) = (1 + 2\varepsilon) - q\left[2 + 2\varepsilon + \frac{\alpha}{q}\right]s \tag{39}$$

$$\lambda = \beta^2 - \alpha\varepsilon - \frac{1}{2}q\left[1 + 2\varepsilon + \frac{\alpha}{q}\right]s \tag{40}$$

Another definition of $\lambda$ at (24),

$$\lambda = \frac{nq}{2}\left[1 + 2\varepsilon + \frac{\alpha}{q}\right] + n(n-1)q \tag{41}$$

and comparing with (12), the exact energy eigenvalues of DKP equation with deformed Hulthen potential are derived as

$$E_{nq} = \frac{V_0}{2q} \pm C\sqrt{\frac{m^2}{4\gamma^2 + C^2} - \frac{V_0^2}{16q^2\gamma^2}} \tag{42}$$

where $C = \sqrt{q^2 - 4\gamma^2} + q(2n+1)$.

Now let us find the corresponding eigenfunctions. If $\phi(s)$ and $y(s)$ are taken as the particular solutions of DKP equation given by (32), the following transformation can be applied as

$$F(s) = \phi(s)y(s) \tag{43}$$

The resulting equation reduces into an equation of hypergeometric type, and the part of $\phi(s)$ is found from the relation $\phi'(s)\phi(s) = \pi(s)\sigma(s)$ as below

$$\phi(s) = s^\varepsilon (1-qs)^{(a+q)2q} \tag{44}$$

The part of $y(s)$ is given by Rodrigues relation,

$$y_{nq}(s) = B_{nq} s^{-2\varepsilon} (1-qs)^{-aq} \frac{d^n}{ds^n}\left[s^{n+2\varepsilon}(1-qs)^{n+(aq)}\right] \tag{45}$$

where $\rho(s) = s^{2\varepsilon}(1-qs)^{aq}$.

The radial wave function $F_n(s)$ is obtained in terms of Jacobi Polynomials at $q \to 1$ limit

$$F_n(s) = C_n s^\varepsilon (1-s)^{(1+b)2} P_n^{(2\varepsilon,b)}(1-2s) \tag{46}$$

where $s = e^{-\delta r}$, $b = \sqrt{1-4\gamma^2}$ and resulting in $y_n(s) \approx P_n^{(2\varepsilon,b)}(1-2s)$.

## 5. Conclusion

In this study, we obtained the energy eigenvalues and eigenfunctions for DKP equation with Hulthen potential by the method of Nikiforov-Uvarov. It is seen that exact solutions of first order Dirac-like DKP equation with deformed Hulthen potential is included the solutions of KG equation for spin zero. Since DKP equation holds for spin zero and one particle unlike KG equation, it is interesting to solve DKP equation in the case of spin one by applying NU method. In addition, energy eigenvalues for different values of deformation parameter q can be calculated and effect of screening parameter $\delta$ on Hulthen potential can be investigated.